\documentclass[pre,aps,twocolumn, showkeys,floatfix,nofootinbib,
superscriptaddress]{revtex4-1}
\usepackage[utf8]{inputenc}
\usepackage{amsmath}
\usepackage{amsfonts}
\usepackage{amssymb}
\usepackage{graphicx}
\usepackage{subfigure}
\usepackage{ulem}
\usepackage{hyperref}
\usepackage{color}
\usepackage{tikz}
\usetikzlibrary{decorations.pathmorphing,patterns,arrows,shapes.misc}
\usepackage{tikz-3dplot}
\usepackage[nomessages]{fp}
\usepackage[first=0, last=360]{lcg}
\usepackage{ifthen}
\usepackage{csvsimple}
\usepackage{comment}
\begin{document}
	
	\title{Visibility graphs of animal foraging trajectories}
	
	\author{Leticia R. Paiva}\email{leticia.paiva@ufsj.edu.br}
	\author{Sidiney G. Alves}\email{sidiney@ufsj.edu.br}
	\affiliation{Departamento de F\'isica e Matem\'atica, Universidade Federal de S\~ao Jo\~ao Del-Rei \\
		36420-000, Ouro Branco, MG, Brazil}
	
	\author{Lucas Lacasa}\email{lucas@ifisc.uib-csic.es}
	\affiliation{Instituto de Fisica Interdisciplinar y Sistemas Complejos (IFISC, UIB-CSIC)}

	\author{Og DeSouza}\email{og.souza@ufv.br}
	\affiliation{Laboratorio de Termitologia, Universidade Federal de Vi\c cosa, 36570-900 Vi\c cosa, Minas Gerais, Brazil}

	\author{Octavio Miramontes}\email{octavio@fisica.unam.mx}
	\affiliation{Departamento de Sistemas Complejos, Instituto de F\'isica, Universidad Nacional
		Autonoma de M\'exico, Ciudad de M\'exico, C.P. 04510, Mexico }
	
	\begin{abstract}

		The study of self-propelled particles is a fast-growing research topic where biologically inspired movement is increasingly becoming of much interest. A relevant example is the collective motion of social insects, whose variety and complexity offer fertile grounds for theoretical abstractions. It has been demonstrated	that the collective motion involved in the searching behavior of termites is consistent with self-similarity, anomalous diffusion and Lévy walks. In this work, we use visibility graphs -- a method that maps time series into graphs and quantifies the signal complexity via graph topological metrics -- in the context of social insects foraging trajectories extracted from experiments. Our analysis indicates that the patterns observed for isolated termites change qualitatively when the termite density is increased, and such change cannot be explained by jamming effects only, pointing to collective effects emerging due to non-trivial foraging interactions between insects as the cause. Moreover, we find that such an onset of complexity is maximized for intermediate termite densities.
	\end{abstract} 
	\maketitle
	
	\section{Introduction}
	
	Animal mobility has attracted a large amount of attention from various disciplines, including biology and physics alike.  Recently, it has surfaced that even Albert Einstein was convinced that the study of animal navigation and movement could inspire the beginning of a new physics~\cite{dyer2021einstein}. Then the 2021 Nobel Prize in physics was awarded for the first time to the study of Complex Systems and was shared by Giorgio Parisi, who has been actively involved in the study of animal collective motion~\cite{ballerini2008interaction, ballerini2008empirical, cavagna2008new, cavagna2010scale, procaccini2011propagating, camperi2012spatially}. Social insects such as ants and termites have been commonly used as conceptual metaphors in statistical physics, for example in the study of percolation, lattice trapping, diffusion \cite{de1980percolation, de2009percolation} and the complexity of self-organization\cite{detrain2006self, miramontes1995order}. Moreover, the study of animal searching and foraging is an emergent field \cite{shlesinger1986levy, Gandhibook} that goes hand in hand with recent and rich developments in the field of random and deterministic walks \cite{metzler2014first,lima2001deterministic,boyer2009levy}. The increasing availability, quality, and quantity of accessible data on animal movement has been useful for answering many questions as well as revealing new ones \cite{Buchanan,Reynolds2010,Wosniack,Reynolds2015,Reynolds2018}. On one hand, it is well established that many animal species in different habitats (terrestrial or marine) search and forage as Lévy walks \cite{Wosniack,Reynolds2015,Reynolds2018,Gandhibook}, which are composed of many short steps alternated with longer steps statistically following power laws \cite{Gandhibook,Reynolds2011}. On the other hand,  when and why organisms perform Lévy walks is still barely known\cite{Reynolds2018}. For example, we yet ignore whether Lévy animal foraging is actually an emergent property \cite{Reynolds2009,Reynolds2013} --certain circumstances induce Lévy patterns-- or Lévy animal foraging strategies are evolutionary adaptations \cite{Wosniack,deJager,Gandhi2008}.
	
	Previous studies suggested that the foraging behavior of both isolated and grouped termites is consistent with anomalous diffusion, in particular Lévy-like walking processes \cite{Miramontes2014,Julieth2022}. In these studies, the termite trajectories were analyzed by means of standard mean squared displacement, structure functions, autocorrelation functions, and probability functions histograms. Although these are valuable, well-established tools to identify normal and anomalous diffusion, it was only recently that more sophisticated studies established that indeed Lévy walks are common in the movements of termites \cite{Paiva2021}. Importantly, time series obtained from biological foraging and searching behavior usually involve nonlinear processes, so their analysis in principle needs the use of different techniques of increasing complexity, e.g. nonlinear correlation functions, and multifractal spectra, which usually require very large data with low experimental noise. Accordingly, new approaches are needed, suitable for the specific type of data at hand. These approaches should naturally incorporate the multiple scales usually found in these processes and be able to filter relevant information from massive data sets which are obtained thanks to technological advances in animal tracking.
	
	Here we suggest that the visibility graph approach \cite{Lacasa2008, Luque2009} --a combinatorial technique whereby time series are mapped into graphs and the analysis of the underlying dynamics is analyzed by quantifying the topological properties-- provides a valuable new angle for the analysis of foraging time series, and bring new insights about their main properties. Accordingly, after mapping experimental foraging data into time series using two alternative protocols, we convert these series into visibility graphs and subsequently study their topological properties for a range of insect densities. 
	Our results first indicate that isolated individuals (with no social contact or external stimuli) perform foraging patterns (innate foraging behavior) which are qualitatively different from the patterns observed when more individuals are confined together. Second, interpretation of these patterns in graph space suggests that the onset of strategies closer to the Levy walk hypothesis are notably enhanced when individuals interact. It seems plausible that this change in movement strategy would be a result of the information exchange and social trapping that is provided by social contacts \cite{Paiva2021}. Third, our visibility-graph analysis suggests the onset of collective effects at intermediate densities and unveils a non-monotonic trend that pinpoints an optimal density where collective effects --and complexity-- are maximized.
	
	The rest of the paper goes as follows. In section II we detail the experimental setup and both the data collection and curation protocols. We also provide details on the methods used for data analysis (visibility graphs) as well as the metrics employed. In section III we provide the results of the experimental foraging patterns in graph space, along with an interpretation of these. In section IV we conclude.

	\section{Experiments and methods}
	
	\subsection{Experimental setup and data collection}

\begin{figure}[ht]
		\begin{center}
			\includegraphics*[width=0.35\textwidth]{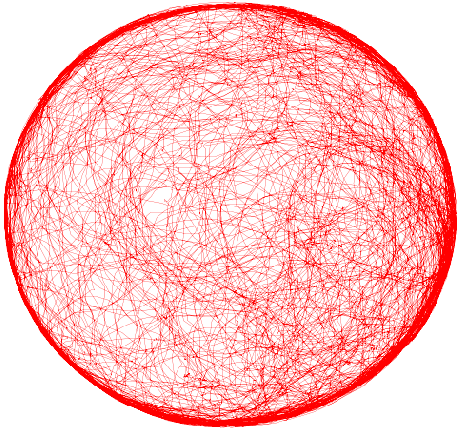}
			\caption{\label{fig:1} Trajectory example of a single termite worker walking and searching in a circular arena given by an empty Petri dish, as reconstructed from the output of an automatized video-tracking device. The individual spends an important amount of time in the border of the arena but sometimes ventures to explore the interior. The result is an intricate web of displacements of different lengths and directions. This is the trajectory no.1 from ref. \cite{Miramontes2014}.}
		\end{center}
		
	\end{figure}

	Termite workers ({\it Cornitermes cumulans}) were collected in Minas Gerais, Brazil, and taken to an acclimatized laboratory as described in detail in \cite{Miramontes2014}. Individuals were ink-marked and placed in circular glass arenas (Petri dishes) under an automatic video recording and tracking system to extract trajectories, see Fig. \ref{fig:1} for an illustration. Time series of positions containing thousands of location points measured each half-second for hours were obtained while the individuals walked through the arenas. Extraction of time series is initially performed using the same methodology as previously reported in \cite{Miramontes2014}.
	
	An important step in the analysis of complex movement data involves the discretization of the continuous movement path of an animal into a series of step lengths. Historically, the first study involving animal tracking and Lévy flights was made on the albatross {\it Diomedea exulans} using dry-wet satellite sensors that did not report actual traveled distances or geopositions, but flight times in-between dry (bird flying) and wet episodes (bird on water). It was suggested initially that these flight times followed power laws\cite{viswanathan1996levy}. With the introduction of GPS and video tracking technologies, it was possible to get coordinates at fixed time frames (the so-called TF method) and a number of studies followed this method \cite{humphries2012foraging, Miramontes2014}.
	However, it is important to observe that, since frames are extracted at fixed times, sequence of displacements from the TF method actually have physical units of distance over time, i.e. this method indeed records average speed sequences.
	An alternative approach to extract actual displacements is given by  Humphries {\it et al.}\cite{Humphries2013}, who proposed a method where turning points are unambiguously identified when the $x$ or $y$-axis movements are computed independently (the $Hx$,  $Hy$ method).
	The time resolution in the TF method is important to both accurately capture fine-grained peculiarities of the foraging trajectory and not oversample it. In this work,
	the steps $\ell_i$ in the TF protocol are defined as successive line segments walked by the termite individuals every $0.5$ second inside the Petri dish. Alternatively, in the Humphries {\it et al.}\cite{Humphries2013} approach, a step is the distance between two adjacent turning points. Both methods have advantages and drawbacks. When the time scale of the animal turns is similar to the tracking device time frame, and the animal velocity is not constant, the TF method gives fair results (such as in termites that have highly tortuous trajectories). However, when the animal travels very long rectilinear distances (as in the albatross) and these are artificially segmented the results may be misleading and the TF method is not a good choice. On the other hand, if the animal is traveling along with unbounded space --such as birds or fish do--, rectilinear distances are well captured with the Humphries {\it et al.} method, but when the animal is caged in small containers, the movements along the borders can be over-represented as a single big step. Here we take an agnostic view and for the sake of comparison and completeness, we will use both methods in parallel. As we will show, the results are qualitatively similar and thus our conclusions are robust against discretization issues.     
	
	\subsection{The visibility algorithm and associated network metrics}
	
	The family of visibility algorithms as graph-theoretic methods for time series were first proposed in \cite{Lacasa2008, Luque2009} and since then they have been successfully applied to a range of problems in many scientific disciplines, from purely theoretical analysis \cite{lacasa2014degree, iacovacci2016sequential, luque2017canonical}, description of nonlinear \cite{lacasa2010description, luque2011feigenbaum, lacasa2018visibility}  and stochastic dynamics \cite{lacasa2009visibility, lacasa2015time}  to applications in physical \cite{elsner2009visibility, murugesan2015combustion, zou2014complex,chowdhuri2021visibility, iacobello2021large}, biological \cite{sannino2017visibility, hasson2018combinatorial, ahmadlou2012improved} and socio-technical systems \cite{flanagan2016irreversibility, Fasmer,Huang,Yang}, even in the arts \cite{gonzalez2020arrow} (see \cite{gao2017complex, Zou2019} for reviews on graph-theoretic methods for signal processing). The methods offer simple and computationally efficient mappings between an ordered sequence --i.e. a time series-- of real-valued data and an associated graph (for the extension to multivariate time series or images see \cite{lacasa2015network, iacovacci2019visibility}).  The structure of the graph inherits properties of the original time series and its underlying dynamical system and provides complementary metrics for signal processing of complex time series.\\
	In this work we consider the original (so-called natural) visibility algorithm. Given a time series ${x_i}_{i=1}^N$ with $x_i = x(t_i)$, the visibility graph is an undirected network of $N$ nodes. Each node is associated to a different datum in the series, such that the graph has an ordered degree sequence. Two nodes $i$ and $j$ are connected in the visibility graph if the time series data $x_i$ and $x_j$ fulfill the following criterion
	\begin{equation}
		\frac{x_i-x_k}{t_k-t_i}> \frac{x_i-x_j}{t_j-t_i}, \ \forall t_k \ \text{with} \ t_i < t_k < t_j
	\end{equation}
	\noindent Therefore, the edges of the network take into account the temporal information explicitly \cite{Zou2019}. For instance, time-series with long-range correlations --such as fractional Brownian motion-- yield visibility graphs where the degree distribution has a power-law tail whose exponent is in turn related to the Hurst exponent of the signal \cite{lacasa2009visibility}. It is also easy to see that extreme events in noisy signals are likely to become hubs in the associated visibility graph, and thus the more prone the signal to have extreme events, the more likely the degree distribution shows fat tails.
	
	To the best of our knowledge, this methodology has not been previously used to analyze the structure of foraging trajectories. By construction, the visibility graph allows to identify those long steps --extreme events in the time series-- related to switching in the foraging strategy (of course, how a step is defined has a crucial role in this analysis; this will be discussed later). Accordingly, a power-law decay on the degree distribution function is a signature of these long jumps in between relatively small steps. 
	
	Once time series are mapped into visibility graphs, we need to consider --among many possibilities-- what specific topological properties of the associated graphs should be retrieved. This selection is mainly driven by interpretability and by previous studies on visibility graphs. We will consider the following topological metrics:
	
	\noindent {\it Mean path length and diameter.} A path in a network is a route that runs along successive links. The path between any two nodes $p$ and $q$ with the smallest number of links is defined as the shortest path $l_{pq}$. By averaging this quantity over all pairs of nodes in the network, we construct the so-called mean path length. Similarly, the diameter $d$ of a network is defined as the largest size of all shortest paths.
	
\noindent {\it Degree distribution and degree entropy.} Recent work has shown that the degree sequence gathers a lot of information from the graph \cite{luque2017canonical}, and thus it is sensible to consider metrics derived from it. In particular, we will consider the degree probability distribution function (PDF) $P(k)$ that quantifies the probability that a randomly chosen node has degree $k$. The degree PDF is just the marginal distribution of the degree sequence and has often been used, in the context of visibility graphs, to explore aspects related to correlations and self-similarity in the signal. We will also consider the entropy of the degree distribution $h=\sum_k P(k) \log P(k)$, or graph entropy, that characterizes the heterogeneity of the graph's wiring architecture. In the case of the horizontal visibility graphs, previous works \cite{luque2017canonical, lacasa2018visibility} strongly suggest that the graph entropy $h$ is indeed related to the Shannon entropy $H$ of its corresponding time series.     
	
	\begin{figure*}[htb]
		\begin{center}
			\includegraphics*[width=7cm]{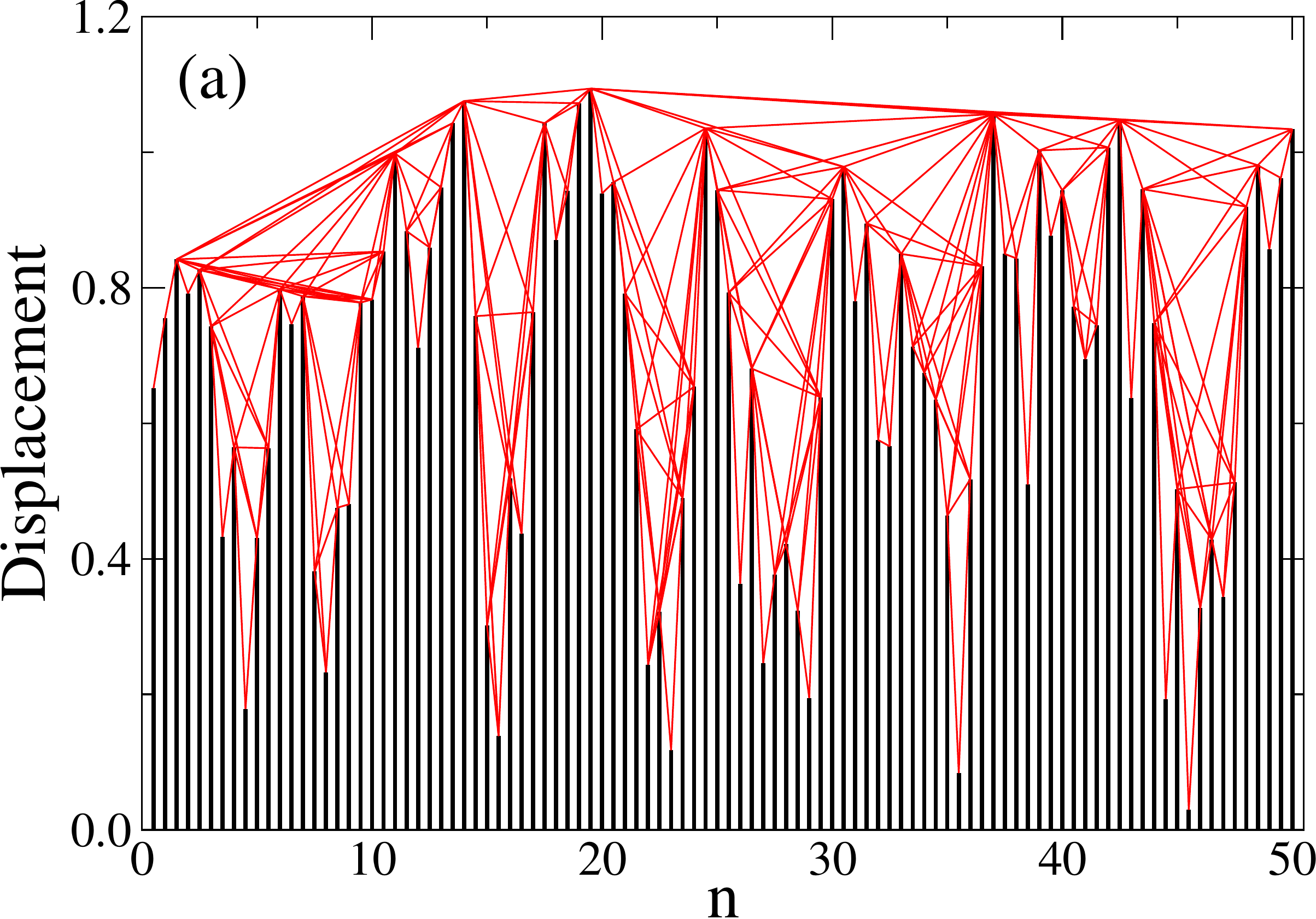}~~~~~~~~
			\includegraphics*[width=7cm]{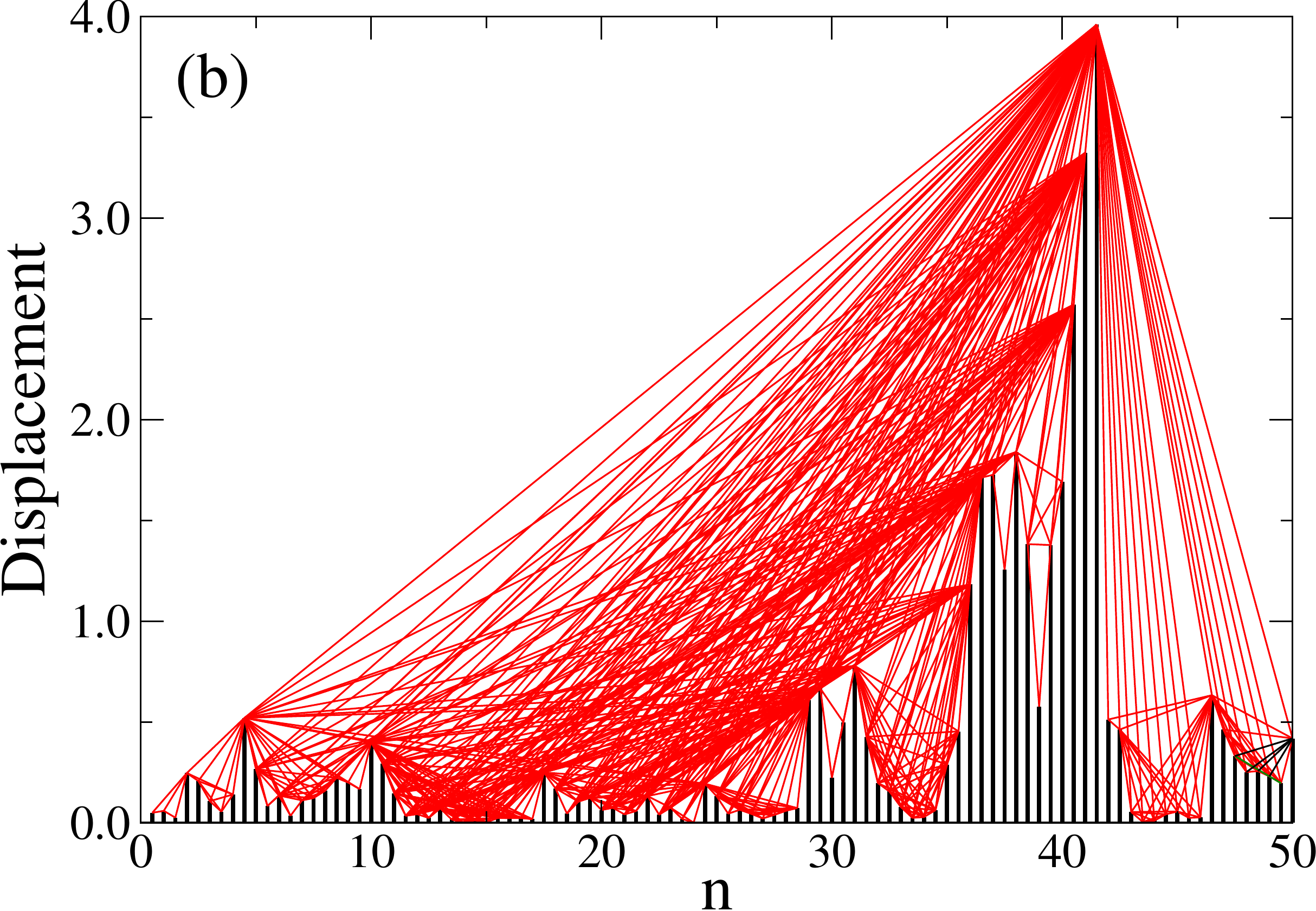}
			\caption{\label{fig:vgraphs} Examples of visibility graphs obtained from 50 steps of termite walking trajectories obtained from experimental arenas (diameter = $5$ cm) containing (a) one isolated individual and (b) eighteen termite workers coming from the same colony. Differences in the graphs are obvious, a target individual in a social context exhibits rare large displacements that results in higher connectivity among the network nodes as compared to the network of an isolated individual. Steps shown here were obtained using the TF method.}
		\end{center}
	\end{figure*}

	\section{Results}
	Using the visibility algorithm \cite{Lacasa2008}, we converted our time series containing termite displacements (with both segmentation methods) into networks and proceeded to compute specific topological properties of these. Following Paiva {\it et al.} \cite{Paiva2021}, we use this new approach to investigate how social interactions between termites modulate their movements and their foraging behavior.
	
	To start, in Fig. \ref{fig:vgraphs} we provide an illustration of an excerpt of the time series of an isolated termite foraging in the arena (panel (a)) and of a termite foraging in a similar arena with higher termite density (18 termites, where the termite under analysis was identified and tracked in the experimental setup). Black, vertical lines denote step length whereas red lines characterize visibility, a pre-processing of the visibility algorithm. It is already easy to see that the foraging pattern is markedly different. We can argue that in the case of a single termite in the arena, since there is no food available neither other termites, the absence of feedback provided by feeding or by social contacts eventually induce a different foraging behavior, in contrast to grouped individuals.
	
	\begin{figure*}[ht]
		\begin{center}
			\includegraphics*[width=8cm]{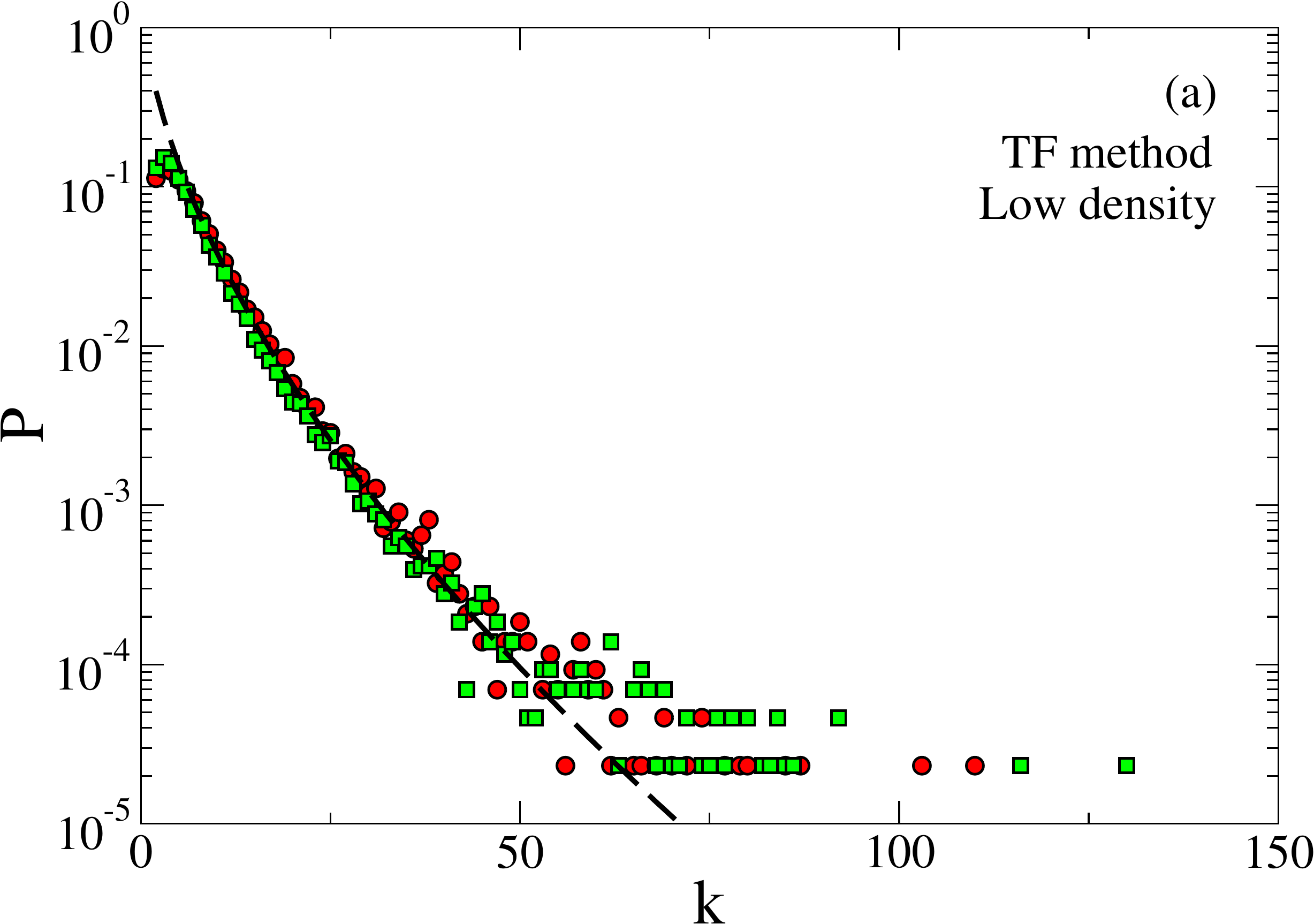}~~~~~~~~\includegraphics*[width=7.6cm]{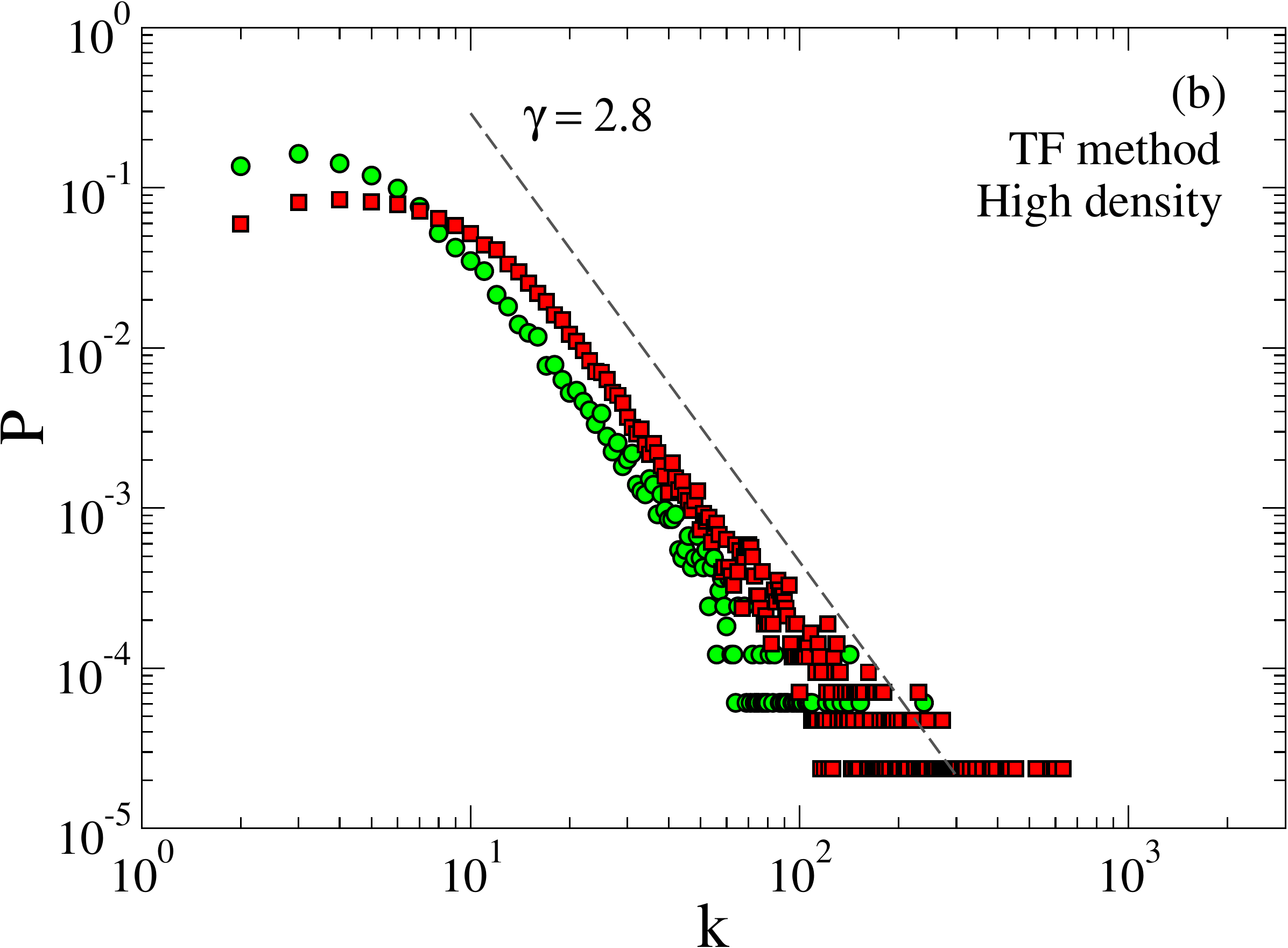}
			\includegraphics*[width=8cm]{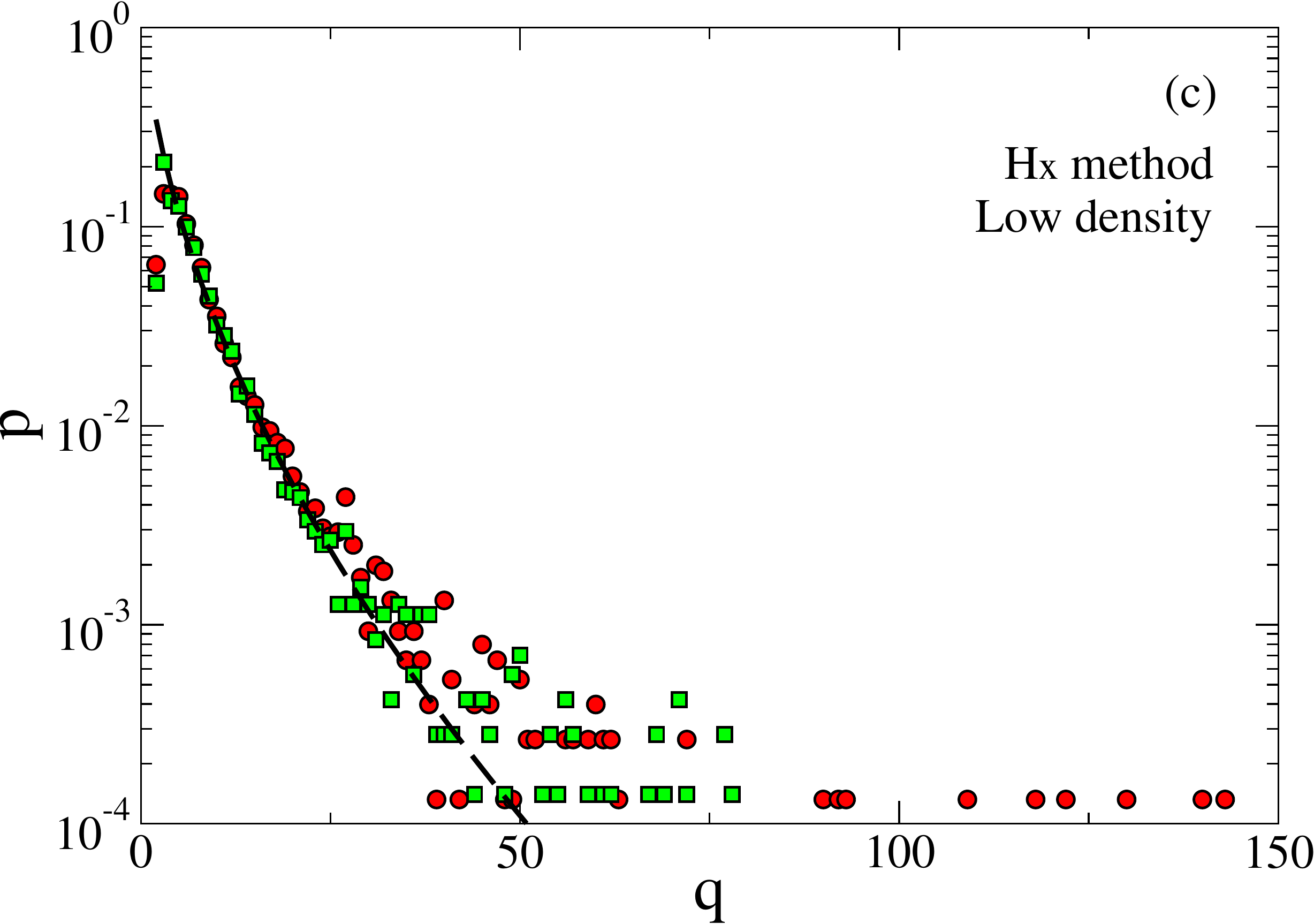}~~~~~~~~\includegraphics*[width=7.6cm]{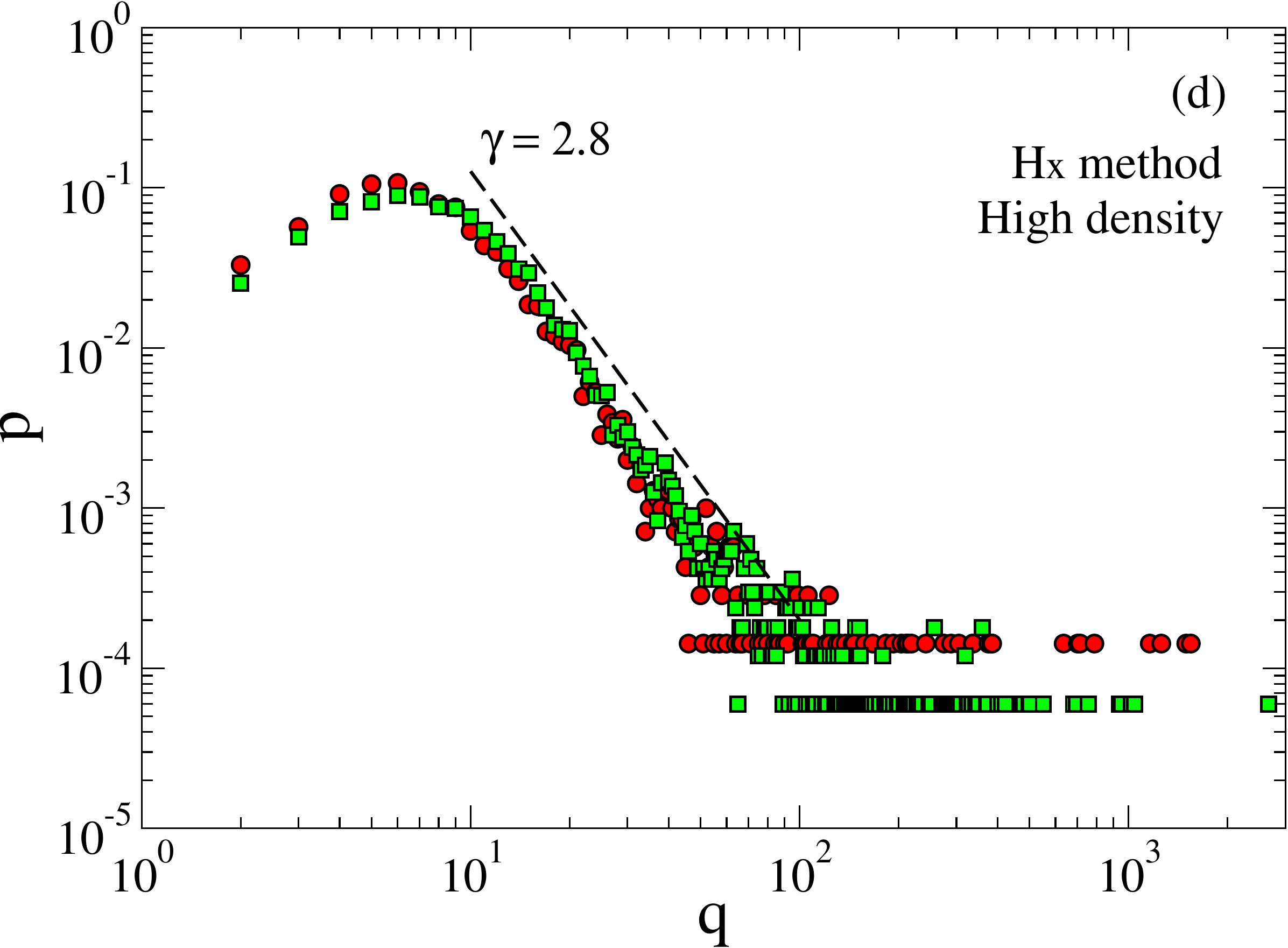}
			\caption{\label{fig:degree}Degree distribution function obtained from the visibility graphs of the trajectories obtained by the TF (a and b) and by the Humphries {\it et al.} method (c and d).  Two examples of distances in low density arenas are depicted in (a and c) where the decay is exponential. Two examples of distances in high density arenas are shown in (b and d), the decay follows a power law with an exponent $\gamma$ close to 2.8. Notice that with both segmentation methods the results are similar. In the Humphries {\it et al.} method, only $Hx$ is shown for simplicity}
		\end{center}
	\end{figure*}
	
	\begin{figure*}[htb]
		\begin{center}
			\includegraphics*[width=7cm]{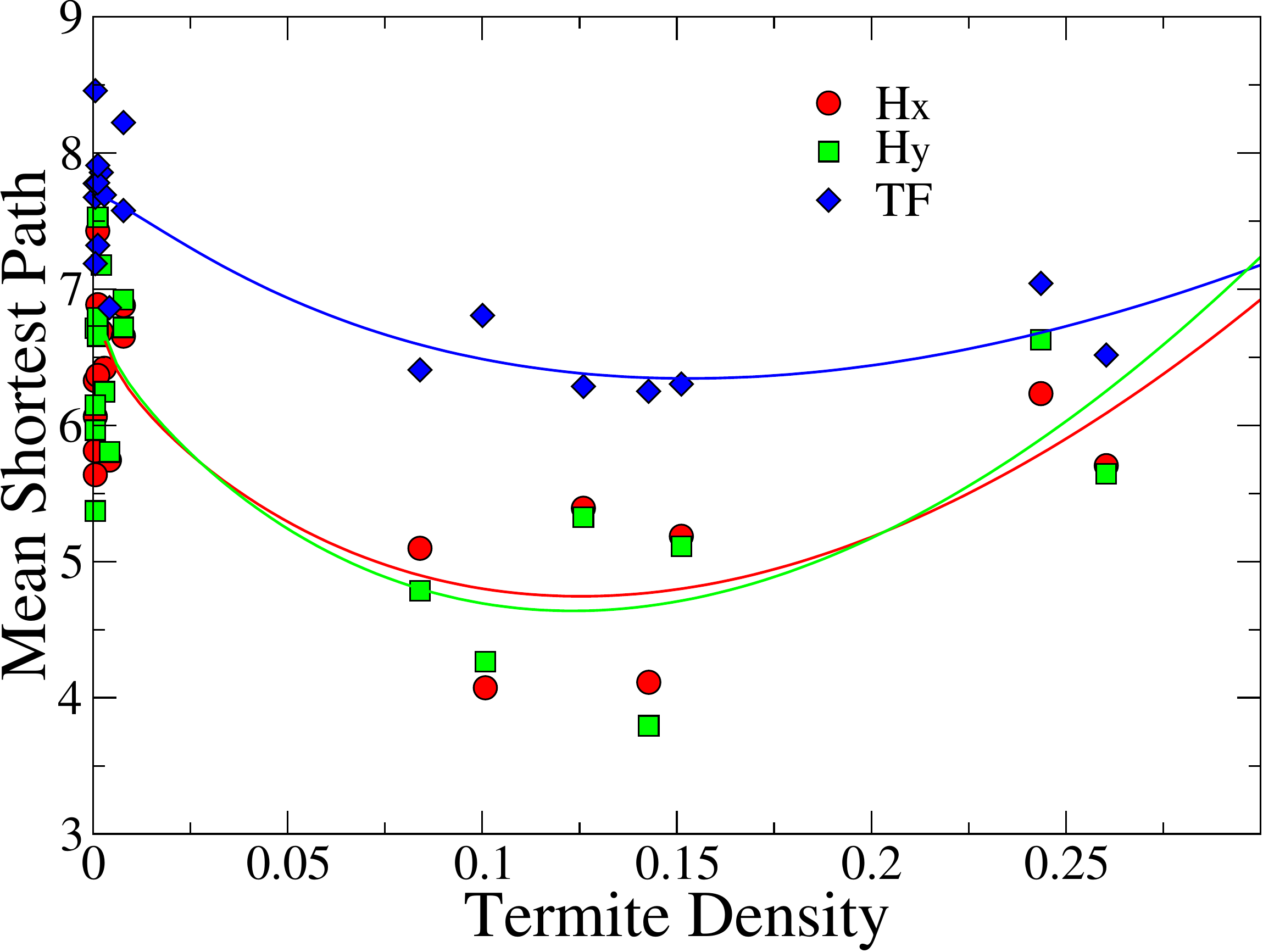}
			\includegraphics*[width=7cm]{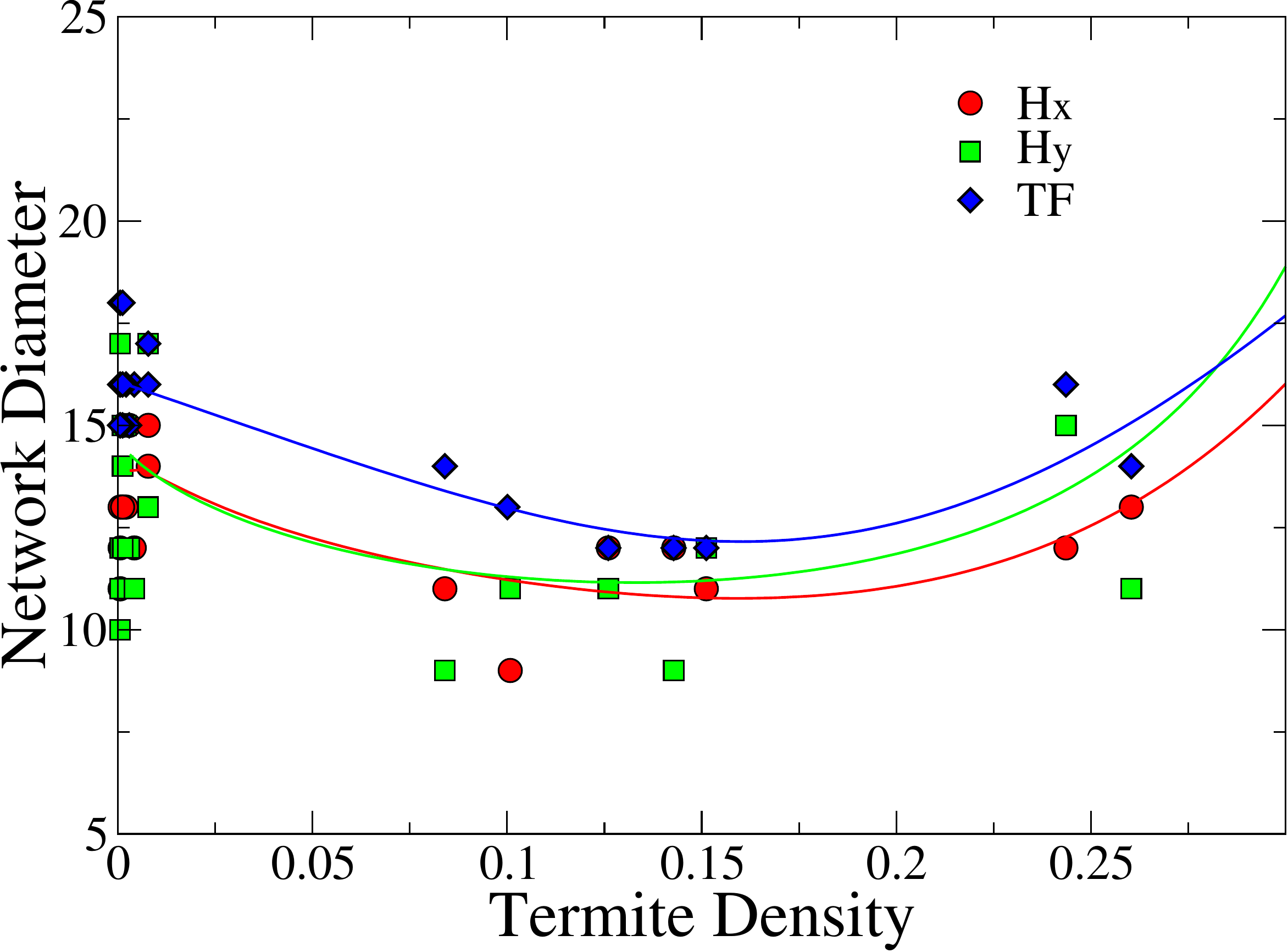}
			\caption{\label{fig:meanpath}
				(Left panel) Mean path density (Mean path length) of the network decreases as the number of termites increases in the arenas. Something similar to the small world phenomenon where increasing disorder produces this trend. The emergence of Levy walks in the grouped termites is the cause here. However, this trend is reversed as the density of termites increases. The minimum was registered at a density of around 0.15.  Both segmentation methods give qualitative similar results. The fitted continuous lines are a guide for the eye only. (Right panel) Network diameter decreases as the number of individuals is increased in the arenas. This parameter is relevant because it quantifies how quickly something (information, diseases, etc.) can spread through the network and also how integrated the network maybe. As in the other cases in the visibility graphs of walking termites, the dynamic behavior of this parameter is caused by the emergence of Lévy walks and the mostly rare long steps that characterize them. Notice however that this trend reaches a minimum value (at around 0.1 -- 0.15) and then the network diameter increases as the density increases.  Both segmentation methods give qualitative similar results. The fitted continuous lines are a guide for the eye only.}
		\end{center}
	\end{figure*}
	
	Second, we now construct the visibility graphs for both isolated termites (several experiments) and for the identified termite in arenas crowded with multiple other termites (several experiments). In Fig. \ref{fig:degree} we plot the degree PDF which describes the heterogeneity of degrees present in the graph (panels (a) and (c) depict isolated termites, panels (b) and (d) depict termites in high termite density arenas, panels (a) and (b) provide results following the TF protocol, and panels (c) and (d) provide results following the Humphries et al. protocol).
	It is well-known that signals with extreme values tend to have visibility graphs with fat-tailed degree PDF, where there is a correlation between extreme events in the signal and hubs in the associated graphs.  Therefore, Lévy-like trajectories naturally translate into visibility graphs with fat-tailed degree distributions. In every case, we find that degree PDF decay slower than an exponential, but with starking differences: whereas isolated termites show degree distributions decaying as a stretched exponential,  
	termites in high-density arenas decay much slower --reaching much larger degrees--, and the tails of these distributions can be fitted to a power-law with robust, non-trivial exponents $\approx 2.8$ for all experiments. This exponent is different from the one found for Brownian motion \cite{lacasa2009visibility} and suggests the onset of long-range temporal correlations. Interpretation of these results is in agreement with by Paiva {\it et al.} \cite{Paiva2021}, where authors argue that Lévy-like trajectories are more likely to appear in termites with social interactions than in isolated ones.
	
	To explore further how visibility graphs `change' as the termite density changes in the experiment, we now consider three different scalar quantities extracted from the visibility graph: the mean path length, the network diameter, and the entropy of the degree distribution (see methods).
	
	The mean path length of the resulting network is plotted as a function of the termite density for all experiments and both segmentation protocols in panel (A) of Fig. \ref{fig:meanpath}, whereas values of the network diameter are plotted in panel (B) of the same figure. In both cases, we observe an initial decreasing trend as the number of termites are increased until a termite density of about 0.15, where this tendency is inverted. 
	The initial decrease is due to the emergence of rare long steps that characterize a Lévy walk in termite groups: these long steps act as degree hubs which act as graph shortcuts and effectively reduce their mean path length and diameter (see next section for interpretation and implications of this finding for the onset of collective behavior).
	
	\begin{figure}[htb]
		\begin{center}
			\includegraphics*[width=7cm]{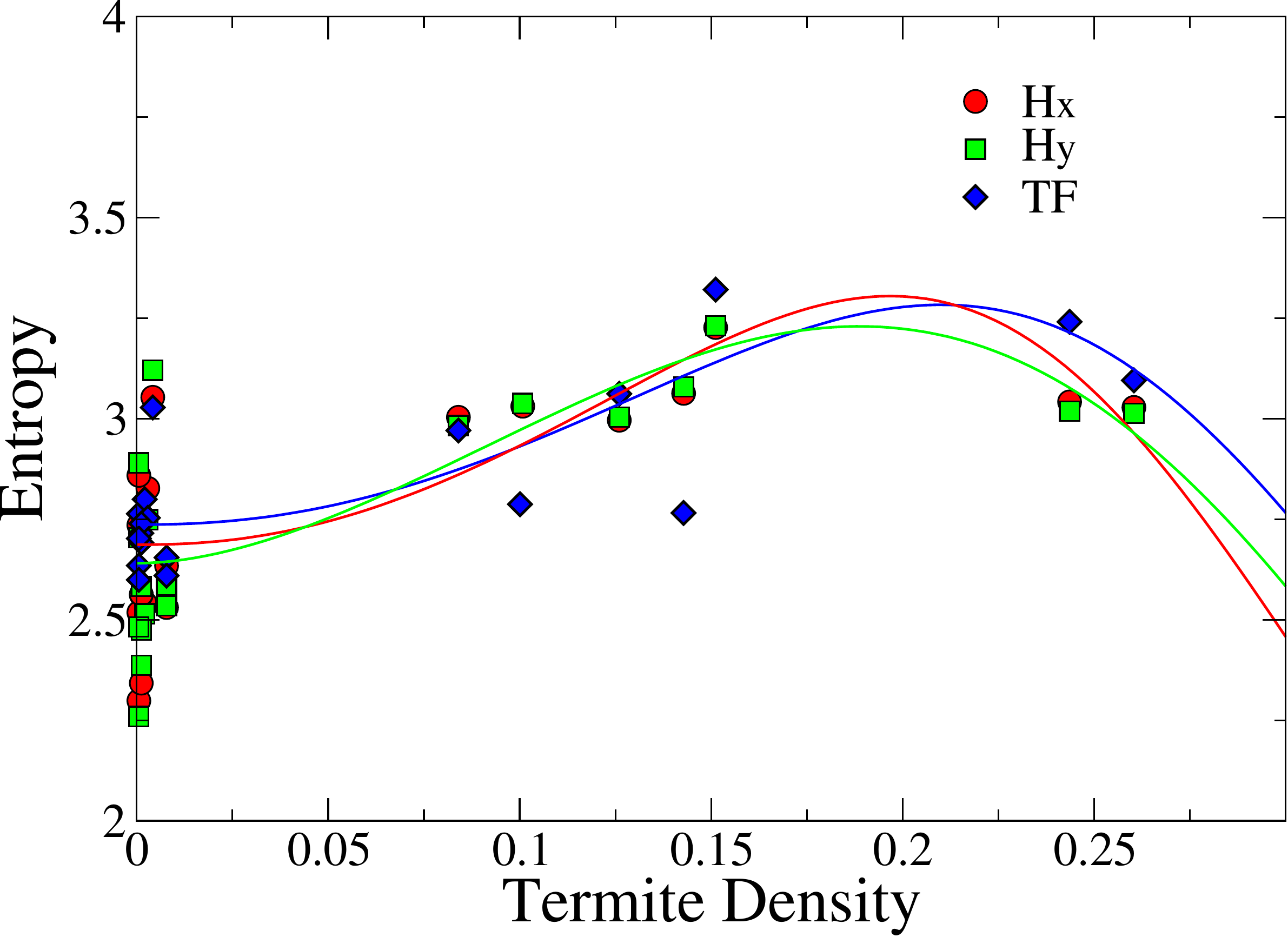}
			\caption{\label{fig:entropy}Graph entropy $S$ versus the number of termites. As the number of insects increases there is a tendency for the entropy to increase signaling more state variability. The increasing trend goes to a peak point and then decreases as the density increases.  Notice that the maximum data was registered at density around 0.15. Graph entropy for both segmentation methods are displayed. The fitted continuous lines are a guide for the eye only.}
		\end{center}
	\end{figure}

	Moving on, in Fig. \ref{fig:entropy} we plot the graph entropy for all experiments and as a function of the termite density, both for the TF protocol (red dots) and the Humphries {\it et al.} protocol (blue and green dots). In both protocols, the graph entropy shows again a non-monotonic behavior as a function of the termite density, initially increasing until reaching a maximum (at the same density where mean path length and diameter are minimal) and then decreasing. Since graph entropy is indeed related to the signal's Shannon entropy, these results suggest that informational content is maximized at intermediate densities. In the next subsection, we wrap up these results and provide an interpretation.
	
	\subsection*{Interpretation: emergent collective effects for intermediate termite density} In this work we were interested in using the metrics depicted above to both quantify foraging patterns and assess whether these change with different termite densities. Our contention is that collective phenomena emerge when termites are grouped together, so
	we therefore need to have an idea of what to expect under a null hypothesis. This subsection is about what to expect if no cooperative phenomenon emerges. Under this premise, the effect of adding termites to the arena and thus increasing its density would only have a size effect. Let us discuss what should be the consequences of this `null model'.
	
	First, when termites are grouped together in the same arena, termites wouldn't be able to reach very large speeds or displacements (they would have more chance of running into another termite, forcing them to decelerate or change course). Accordingly, the marginal distribution $P(\delta)$ should --under the null model-- have shorter tails for higher densities, as extreme speeds or extreme long displacements couldn't be reached due to jamming. This, however, does not happen, in fact, the opposite is true.
	
	Second, a similar argument would suggest that hubs in visibility graphs should be rarer as termite density increases. Since the onset of hubs tend to reduce both the graph mean path length and the network diameter, then the null model would predict that in the visibility graph these two metrics should smoothly {\it increase} as the termite density increases. This effect is observed only in the limit of large density --high jamming--,  but for an intermediate range the opposite effect is found, which leads us to conclude that in such a range collective effects do emerge and qualitatively change the termite foraging pattern. These effects reach a maximum density of around 0.15, and subsequently vanish due to strong jamming effects. 
	
	Third, if the only effect of a high termite density was jamming (i.e. an effectively reduced space), then the time series would be expected to display a higher roughness for higher density. According to previous work, this indeed would imply that the degree distribution should be more and more homogeneous --i.e. have a lower entropy-- as termite density increases. This decreasing effect is indeed found at large termite densities, but, interestingly, again for intermediate densities, the effect is precisely the opposite. This is an additional evidence that suggests that in such a range of densities the null model hypothesis does not hold and increasing termite density boosts nontrivial collective effects. Such collective effects attain a maximum at an intermediate density of about 0.15, beyond which jamming breaks down coherence.
	
	\section{Conclusions}
	
	Animal movement patterns, searching and foraging behavior and, in general, the behavior of self-propelled particles, have been research topics of increasing interest. Individual or collective motion are currently very active subjects in the study of complex systems and statistical physics. New methodologies are continuously devised to help explain their many aspects. One of them, the visibility graph algorithm is a novel approach in network analysis that is useful to identify additional patterns of dynamical behavior to those found in the study of their corresponding time series \cite{Lacasa2008}.
	
	By mapping time series into networks, anomalous diffusion in animal foraging and animal movement can be examined in a broader way. We have examined time series containing termite movements while they explore empty arenas and have recorded the time series of individuals under isolation and in groups of different sizes. Our subsequent analysis using visibility graphs suggests that there is a qualitative change in the foraging behavior between isolated and grouped termites, where the latter display more complex patterns, closer to Lévy walk phenomenology. Such change is not due to a trivial size effect and can be argued to be reminiscent of the spontaneous onset of collective behavior, driven by social interactions. Interestingly, we find that such complexity attains a maximum at an intermediate termite density.
	
	Finally, it is important to observe that some authors have suggested there exists a relation between termite group size and quantitative biological relevant phenomenology such as survival, stress and starvation resistance, and tolerance to disease and poisoning~\cite{desouza2004non, desouza2001social, miramontes2008individual}. Those studies concluded that social facilitation was the mechanism that helps attain maximum peaks of these quantities at low densities. Our results are well aligned to these works, as our visibility graph analysis suggests that the emergence of Lévy walks is indeed a density-induced collective phenomenon. We wonder if, at least in the case of social insects, Lévy-like processes might thus emerge not only for foraging success and optimized biological encounters~\cite{Paiva2021}, but also perhaps driven by other biological mechanisms such as those mentioned above.  
	
	\section{Acknowledgements}
	
	We very much appreciate the stimulating comments by Dr. Bartolome Luque. This study was financed in part by the Coordenação de Aperfeiçoamento de Pessoal de Nível Superior – Brasil (CAPES) – Finance Code 001, as well as the Minas Gerais State Foundation for the Support of Scientific Research (Fapemig), and the Brazilian National Council for Scientific Development (CNPq). We are grateful to Prof. Eraldo Lima for granting access to Ethovision and the facilities of the Semiochemicals lab at Federal Univ. of Viçosa. ODS holds CNPq Fellowship \# PQ 307990/2017-6. 
	SGA holds CNPq Fellowship \# PQ 306778/2015-7. OM acknowledges financial support from a UNAM-PAPIIT grant \# IN107619 and a CIENCIA SEM FRONTEIRAS grant from CAPES-Brazil (2013-2015). OM also thanks the Lab of Termitology at UFV-Brazil for their hospitality during multiple research visits to them. LL acknowledges funding from project DYNDEEP (EUR2021-122007) and project MISLAND (PID2020-114324GB-C22), both projects funded by Spanish Ministry of Science and Innovation. We also thank the free software community for the computational applications needed for data storage and manipulation, data analyses, image processing, typesetting, etc., through GNU-Linux/Debian, Ubuntu, \LaTeX, BibTeX, Python, Grace, Overleaf, among others. This is contribution \# 84 from the Lab of Termitology at UFV (\url{http://www.isoptera.ufv.br}).

	\bibliography{visible_termites}
\end{document}